# DRAFT ARTICLE
for IEEE Spectrum

Original working title:

# Back to the Future:
## The Case for Reversible Computing

**Michael P. Frank**[*]

Version 5.7 (ArXiv Preprint v2) as of 3/7/18

*Abstract: There is one, and only one way, consistent with fundamental physics, that the efficiency of general digital computation can continue increasing indefinitely, and that is to apply the principles of reversible computing. We need to begin intensive development work on this technology soon if we want to maintain advances in computing and the attendant economic growth*

***NOTE:*** *This paper is an extended author's preprint of the feature article titled "Throwing Computing Into Reverse"[1] (print) or "The Future of Computing Depends on Making it Reversible"[2] (online), published by IEEE Spectrum in Aug.-Sep. 2017. This preprint is based on the original draft manuscript that the author submitted to Spectrum, prior to IEEE edits and feedback from external readers.*

**Since the dawn of the transistor**, technologists, and the world at large, have grown accustomed to a steady trend of exponentially-improving performance for information technologies at any given cost level. This performance growth has been enabled by the underlying trend, described by Moore's Law,[3] of the exponentially-increasing number of electronic devices (such as transistors) that can be fabricated on an integrated circuit. According to the classic rules of semiconductor scaling,[4] as transistors were made smaller, they became simultaneously cheaper, faster, and more energy-efficient, a massive win-win-win scenario, which resulted in concordantly massive investments in the ongoing push to advance semiconductor fabrication technology to ever-smaller length scales.

Unfortunately, there is today a growing consensus within industry, academia, and government labs that semiconductor scaling has not very much life left; maybe 10 years or so, at best. Multiple issues that come into play as we dive deeper into the nanoscale mean that the classic scaling trends are losing steam. Already, the decreasing logic voltages required due to various short-channel effects resulted in the plateauing of clock speeds more than a decade ago, driving the shift towards today's multi-core architectures. But now, even multi-core architectures face the looming threat of increasing amounts of "dark silicon,"[5,6] as heat dissipation constraints prevent us from being able to cram any more operations per second into each unit of chip area, due to the energy that is converted to heat in each operation. Fundamentally, achieving higher performance within a system of any given size, cost, and power budget requires that individual operations have to become more energy-efficient, and the energy efficiency of conventional digital semiconductor technology is beginning to plateau for a variety of reasons, all of which can ultimately be traced back to fundamental physical issues. Looking forward, as transistors become smaller, their per-area leakage current and standby power increases; meanwhile, as signal energies are decreased, thermal fluctuations become more significant, eventually preventing any further progress[7] within the traditional computing paradigm. Heroic efforts are being made within the semiconductor industry to try to allay and forestall these problems, but the solutions are becoming ever more expensive to deploy, with new leading-edge chip fabrication plants ("fabs") now costing on the order of $10 billion each.[8] But, it's worth pointing out that no level of spending can ever defeat the laws of physics. Beyond some point that is, now, not very far away, a new conventionally-designed computer that simply has smaller transistors would no longer be any cheaper, faster, or more energy-efficient than its predecessors, and at that point, the progress of conventional semiconductor technology will stop, being no longer economically justifiable. The writing is on the wall.

[*] This work was supported by the Laboratory Directed Research and Development program at Sandia National Laboratories and by the Advanced Simulation and Computing program under the U.S. Department of Energy's National Nuclear Security Administration (NNSA). Sandia National Laboratories is a multimission laboratory managed and operated by National Technology and Engineering Solutions of Sandia, LLC., a wholly owned subsidiary of Honeywell International, Inc., for NNSA under contract DE-NA0003525. Approved for public release SAND2017-9424 O.

Obviously, however, we would prefer if the progress in the cost-efficiency of information technology were *not* to stop, since a large portion of our potential future economic progress would be empowered by the continuing advancement of this technology. So then the question arises, can we perhaps keep progress in computing going by transitioning over to some new technology base that is *not* "conventional semiconductor technology?"

Unfortunately, some of the most crucial fundamental physical barriers that will prevent conventional complementary metal-oxide-semiconductor (CMOS) technology from advancing very much further will also still apply, in a more or less comparable way, to any alternative technology as well, as long as we insist on maintaining the present-day computing paradigm, namely *irreversible computing*. No other *irreversible* "beyond CMOS" technology can ever be very much better than end-of-the-line CMOS—at most, it will be better only by some relatively modest, limited factor. However, for several decades now, we have known that there exists a theoretically possible alternative computing paradigm, called *reversible computing*. Developing reversible computing (and then continuing to improve it) is in fact the *only possible way,* within the laws of physics, that we might be able to keep computer energy-efficiency and cost-efficiency for general applications increasing indefinitely, far into the future.

So far, the concept of reversible computing has not received very much attention, which has perhaps made sense up until now, since it is indeed highly challenging to implement effectively, and the alternative of advancing conventional technology was much easier. Nevertheless, significant conceptual progress on reversible computing has been made over the decades by the small number of researchers pursuing it. Still, many difficult problems[9] remain to be solved, and it is going to require a much larger effort, looking forwards, to address them. But, this effort will be highly worthwhile, because the potential upside that reversible computing offers is *many orders of magnitude* of information technology efficiency improvements, with *associated economic advancements*, compared to all possible irreversible computing technologies. With the end of conventional technology now in sight, it's now time that the world's best physics and engineering minds turn committed attention towards reversible computing, and begin an all-out effort to tackle its remaining engineering challenges, so as to bring this idea to practical fruition.

**The first person** to describe the energy-efficiency implications of the conventional irreversible computing paradigm was Rolf Landauer of IBM, who wrote a paper in 1961 called "Irreversibility and Heat Generation in the Computing Process."[10] This paper has generated controversy in some circles, but Landauer's key insight in this paper really does just follow directly as an immediate logical consequence of our most thorough, battle-tested understanding of fundamental physics. All of our most fundamental laws of low-level physical dynamics are *reversible*, meaning that if you were to have complete knowledge of the state of any given closed system at some time, and of the values of all of the relevant physical constants, you could always, conceptually, run the laws of physics backwards, and determine the system's past state at any previous time exactly. (This is even true in quantum mechanics, if you knew the exact quantum state of the system.) As a consequence, it is impossible to have a situation wherein two different possible detailed states at some earlier time, could both evolve to become the exact *same* detailed state as each other at some later time, since this would mean that the earlier state couldn't be uniquely determined from the later one. In other words, at the lowest level in physics, *information cannot be destroyed*. It's important to realize how absolutely essential to our most basic understanding of physics this principle is. If it wasn't true, then the Second Law of Thermodynamics (which says that entropy cannot decrease) could not be true, since entropy is just unknown information. If physics was not reversible, then entropy could simply vanish, and the Second Law would not hold.

How does the indestructibility of information relate to the energy efficiency of irreversible computing? The point is that, since physics is reversible, whenever we think that we are destroying some information in a computer, we actually are not. Putatively "irreversible" operations (such as erasing a bit of information, or destructively overwriting it with a newly-computed value) are, in some sense, really just a convenient fiction. What's actually happening, at the most fundamental level, is that the physical information that is embodied within the systems whose state we think we are "erasing" or "overwriting" (*e.g.*, a circuit node charged to a particular voltage) is simply getting pushed out into the machine's thermal environment, where it effectively becomes entropy (in essence, randomized information), and is manifested as heat. To increase the entropy of a thermal environment at temperature $T$ by an increment $\Delta S$ requires adding an increment of heat $\Delta Q = T\Delta S$ to that environment; that is simply the thermodynamic definition of temperature. Therefore, doing anything that is *logically irreversible*, *i.e.*, that "loses information" in a computer, implies converting useful energy into the less-useful form of increased heat in the environment. An irreversible computer (which all conventional computers are) can therefore be thought of as being essentially just a kind of expensive electric heater that happens to also perform a small amount of computation, very inefficiently, as a side effect—it is, in a sense, maximally inefficient, since each increment of energy that ends up getting thermalized by the system only gets used to represent a single digital bit's worth of computed information for a short time, until that bit gets erased (by grounding a circuit node, say), at which point all of the stored energy associated with that bit gets degraded into heat.

**Can we do better?** Landauer began to consider this question in his 1961 paper. He noted that one can also consider *logically reversible* computational operations, meaning ones that do not attempt to merge together any possible computational states, but which instead only transform them, one to one. Such operations could, in principle, be carried out in a thermodynamically reversible way that produced no entropy, in which case any energy associated with the information-bearing signals in the system would *not* necessarily have to be dissipated to heat, but could, instead, potentially be reused for subsequent operations. Landauer also noted that furthermore, any desired logically irreversible computational operation could, indeed, be embedded into a reversible one, by simply setting aside any information that was no longer needed, rather than erasing it right away. However, Landauer originally thought, at the time, that doing this was only delaying the inevitable, as the information would still need to be erased eventually, when the available memory filled up.

It was left to Landauer's younger colleague, Charles Bennett, to point out[11] that instead, one could reversibly *decompute* intermediate results after a desired result was produced in order to recover the temporary memory used, so that it could be reused for subsequent computations, without ever having needed to erase or irreversibly overwrite it. In this way, reversible computations, if implemented with nearly thermodynamically reversible hardware, could in principle circumvent Landauer's limit on energy dissipation, something that traditional irreversible computations could never do. Over time, more memory-efficient reversible algorithms were developed by Bennett[12] and others,[13] and Bennett himself went on to make pioneering contributions[14,15] to the new field of quantum information and quantum computation[16] that emerged in the late 20th century, a field in which reversible algorithms were essential to its promise of obtaining exponential quantum speedups[17] for certain specialized problems.

But meanwhile, Bennett's original vision, of using reversible computing principles to make *all* computation much more energy-efficient, languished in the academic backwaters. In a nutshell, the problem was simply that, it is *really hard* to actually engineer a system that does something computationally interesting without producing a significant amount of entropy with each operation, despite the fact that, to date, we know of no valid argument from fundamental physics showing that it is impossible in general to approach perfect thermodynamic reversibility in appropriately-designed computational mechanisms, as technology improves.

Indeed, one can say that, at its very foundation, *physics itself*, as a deterministic dynamical system, *is already exactly an example* of a reversible computational process that produces no entropy, if it is viewed from a hypothetical omniscient perspective that tracks the exact quantum state of the universe. Physical time-evolution can itself be viewed as simply being a computation that takes the "old state" of the universe, and computes the "new state" from it, in place, in a one-to-one fashion; and no new uncertainty is introduced during this process, if you hypothetically knew the exact laws of physics and followed the evolution exactly.[†] All of the entropy that appears to exist in the universe can be considered to be simply an illusion that is only suffered by limited beings such as ourselves because we lack such an omniscient perspective.[‡] For this reason, it seems quite likely that reversible computing *can never be proven to be impossible*, since this would require proving that the omniscient perspective couldn't exist, even hypothetically, which would be an odd thing to try to prove, given that physicists assume the existence of such a perspective all the time.

If we accept this basic argument as to why reversible computing must, in principle, be possible, the question then becomes simply, how closely can we manage to arrange for a well-engineered piece of the universe to approach the reversible ideal in practice, given that our knowledge of real systems, and the laws of physics, is more limited? Exactly how many computationally-useful transformations can we arrange for a suitably-engineered system to go through, reusing its active energy repeatedly over many computational steps, before we lose track of its state, and its active energy gets dissipated to heat? Despite numerous attempts to definitively answer this question, we still know of no valid reasons from fundamental physics why, with increasingly sophisticated engineering, the energy efficiency of reversible computations cannot be made indefinitely large over time, as technology improves.

**The first detailed attempt**, following Bennett's early conceptual descriptions, to give a physical picture of how to actually implement reversible computing in a reasonably effective way was carried out by Ed Fredkin (an early director of the MIT Laboratory for Computer Science) together with his colleague Tommaso Toffoli, in their Information Mechanics research group. Fredkin and Toffoli's theoretical work focused on what I call *unconditionally reversible* computational operations. (This is *not* the most general model of reversible operations that avoid Landauer's limit, but we'll get to that later.) Fredkin and Toffoli proposed that such operations could, in principle, be carried out by systems such as idealized electronic circuits[20] that used inductors to shuttle charge packets back and forth between capacitors, or, in the mechanical domain, by ideal rigid spheres[21] bouncing off of each other and fixed barriers in narrowly-constrained trajectories. Unfortunately, these early visions were too idealized to be immediately realized in practice, but the abstract reversible logic model developed in the course of this research, involving computational primitives such as the ones now known as the *Fredkin gate*[22] and the *Toffoli* gate,[23] became the foundation of much of the subsequent theoretical work in reversible computing, including, eventually, the model of quantum logic networks,[24] developed later by David Deutsch.

---

[†] It's a mathematical fact of quantum theory that the Von Neumann entropy[18] of any mixed state is constant under any definite unitary time-evolution operator, such as that induced by any specific field-theoretic Lagrangian.

[‡] Mathematically, even for pure states, tracing out unobserved remote subsystems to produce a reduced density matrix description of a local system results in subjective decoherence,[19] and an increase in apparent entropy.

In the meantime, other researchers continued to explore possible electronic implementations. Researchers at places such as Caltech,[25] USC,[26] Rutgers,[27] and Xerox PARC[28,29] developed early concepts for *adiabatic circuits* which, along the lines of Fredkin and Toffoli's previous proposal, transferred charge in a gradual, controlled way between circuit nodes, dissipating, in principle, only a small fraction of the signal energy with each transition. These efforts led back around to find fertile ground once again at MIT, where members of the Information Mechanics group such as cellular automata expert[30] Norman Margolus still resided. Margolus had also developed some of the early quantum-mechanical models[31,32] of reversible computing, building on previous concepts by Richard Feynman.[33,34] I joined this group in 1996, after having previously found Bennett's ideas to be quite useful when I was working with reversible thermochemical processes in the context of DNA computing. Under the sage guidance of legendary hacker[35] Tom Knight (who subsequently pioneered the field of synthetic biology[36]), his student, Saed Younis, had shown for the first time[37] that adiabatic circuits could implement arbitrary pipelined, sequential reversible logic. Subsequent students including Carlin Vieri[38] and I built on that foundation to design fully-reversible processors of various types,[39,40] as a proof-of-concept establishing that there were, indeed, no fundamental barriers preventing the entire discipline of computer architecture from being translated over to the reversible realm.

Meanwhile, other researchers had explored alternative approaches to implementing reversible computing that were not based on semiconductor electronics at all. The early nanotechnology visionary[41] K. Erik Drexler (whose first course in that subject I had taken at Stanford in 1988) produced detailed designs[42,43] for reversible nanomechanical logic devices made from nanostructured diamondoid materials, which could theoretically be assembled whenever Drexler's vision of general-purpose molecular nanofabrication technology finally came to fruition. (Many people later read Neal Stephenson's science fiction book *The Diamond Age*,[44] which was inspired by Drexler's vision.) Also, over the decades, Russian and Japanese researchers had been developing reversible superconducting electronic devices, such as the similarly-named but distinct *parametric quantron*[45] and *quantum flux parametron*.[46] And a group at Notre Dame[47] was studying how to do logic using adiabatic transformations of interacting single electrons in arrays of quantum dots. To those of us who were working on reversible computing in the 1990s, it seemed that, based on the wide range of promising implementation concepts such as these that had already been proposed, some kind of practical reversible computing technology might not be very far away.

**What happened?** What happened was simply that conventional irreversible semiconductor technology continued improving rapidly through the 1990s and early 2000s. This presented an enormous uphill battle that any radical new alternative computing technology would need to overcome. To succeed, it would not only have to replicate a large part of the effort that had already been invested in developing the entire industrial infrastructure of semiconductor fabrication equipment, design tools, and the associated engineering workforce development, but it would also have to provide a solution that could be competitive not only against contemporaneous conventional technology, but also against *all of the future generations of conventional technology that would subsequently become available before the new technology solutions would be ready*. In other words, the time horizon before this line of research could realistically even hope to begin to pay off in practice was more or less doomed to not occur any time before traditional irreversible technology ran out of steam.

As a result, there was very little will among funding agencies, in the early 2000s, to invest in this area. Young researchers like myself found ourselves caught in a "valley of death" between some proposal reviewers at pure science agencies like NSF who saw the research as "too practical" in focus, and thought it should be therefore be funded by industry, versus industry funding sources who, for the most part, found it to be a radical idea well outside the scope of research they would support at a significant level. Furthermore, all of this general reticence was compounded by an extreme lack of familiarity with the field among computing professionals, along with many widespread misconceptions about it. In my experience, in these years there were only a few lucky breaks: *E.g.*, a far-sighted program manager in the DARPA Scalable Computing Systems program funded our group at MIT for a few years, and the industry-backed Semiconductor Research Corporation supported a portion of my subsequent work at the University of Florida for a brief period under their Cross-Disciplinary Semiconductor Research program. But for the most part, government and private funding sources held back from committing any major investments to reversible computing, and so the field mostly languished.

**Nevertheless, some progress was made**. While teaching at UF in the early 2000s, I fleshed out some of my earlier theoretical work on the scalability[48] of reversible architectures in more depth,[49,50,51] and also clearly defined the requirements[52] for circuits to be *truly, fully* adiabatic—most of the "adiabatic circuit" designs in the literature actually don't meet these requirements, and so are significantly less energy-efficient than more suitably-designed adiabatic circuits can be. I developed a simplified new truly/fully-adiabatic logic family called 2LAL[53] (two-level adiabatic logic, see Box 2). A student, Krishna Natarajan, and I showed in detailed simulations that 2LAL could dissipate as little as 1 electron-volt of energy[54] per transistor per cycle—which was only about 0.001% of the energy of the logic signals in the CMOS technology we were using. Meanwhile, I began showing, in my lectures[55] in my *Physical Limits of Computing* course and elsewhere,[56] how the classic Fredkin-Toffoli model of reversible computing could be significantly generalized, to what I have sometimes called *conditionally reversible*[57] computing. A better name for it might be simply, generalized reversible computing[58] (GRC). The key observation in GRC is that initial states that are simply *disallowed* (that is, that are arranged to occur only with probability zero) *do not contribute at all* to the amount of information that is lost in an operation. Therefore, for a computational operation to avoid ejecting entropy to the environment, it does *not* have to avoid state mergers for *all* of the initial digital states that are combinatorially possible, but *only* for the *subset* of the initial states that are *actually allowed* in a given design (see Box 1). This observation greatly enlarges the set of computational operations that can be seen to be reversible in appropriate operating contexts, and in fact, it is essential for properly understanding the connection between reversible computing theory, and what the adiabatic circuit implementations (which had already existed for some time) were actually doing.

Many blanket statements that are frequently made about reversible computing in the context of the classic Fredkin-Toffoli model aren't actually true in this more general picture. For example, one often sees the statement that "conventional two-input Boolean logic gates such as AND and OR are fundamentally irreversible." But in fact, that is *only* the case for gates that *actively consume* their inputs, which real-world CMOS gates never do, or that *destructively overwrite* their outputs, which conventional CMOS gates, including single-input NOT gates, *always* do, but which is not strictly necessary. Even Landauer's original paper[10] had already pointed out that ordinary Boolean AND and OR operations can be embedded within reversible operations, and many of the adiabatic logic styles[37,53] as well as the alternative implementations such as rod logic that were invented[42,43] took advantage of the fact that OR or AND gates can operate by simply reversibly *transforming* their output nodes from some predetermined initial value, to a new value that is in the process of being computed, without suffering any inconsistency whatsoever with Landauer's model, which is properly understood to apply *only* to the *transformational* relation holding between the *old* and *new* states of the system, but *not* necessarily to the *functional* relationship between the logical input and a computed output. (See Box 1.)

But, unfortunately, the critically important *physical* distinction between the different roles played by the old/new versus input/output dichotomies ended up becoming blurred in the minds of many of the researchers who were exposed to the Fredkin-Toffoli model, which applied these dichotomies only within a restricted theoretical context in which they happened to be equivalent. As a result, many of the subsequent researchers who came into the field never quite properly understood the old/new versus input/output distinction, and repeated the resulting misconceptions, confusing others. One still periodically sees papers published[59] by researchers who, unfamiliar with the full depth of the prior art, believe that they are newly discovering the fact that AND and OR operations that do not consume their inputs or destructively overwrite their outputs can indeed be carried out reversibly without any entropy increase, and mistakenly conclude from this that the connection between logical and physical reversibility, which is the entire rationale for the field of reversible computing, is in error.

In fact, the rationale for reversible computing is perfectly valid; it's only the widespread misconception about *what logically reversible computational operations really are* that is mistaken. The most general class of reversible operations encompassed within GRC is somewhat broader than the class described by Fredkin and Toffoli, which is restricted to a theoretical context in which all input signals are consumed or transformed in-place, and wherein all possible input combinations are considered likely to occur. But more generally, *any* computational operation that does not expel any of the entropy that is contained in a given initial-state probability distribution is, by definition, logically reversible. This includes the more generalized class of conditionally-reversible operations, such as the reversible OR in Box 1, whenever their preconditions have probability zero of being violated.[58] But in any case, all logically reversible computations, since they cannot lose any information, must still always be designed with attention being paid at all times to where all of the information embedded in the computation is going. That essential new design constraint is why the field of reversible computing is unavoidably needed, and why we must eventually begin more widespread efforts to develop (and figure out practical, efficient ways to implement) reversible logic architectures[60] and algorithms.[61]

Those of us who have been most deeply immersed in the foundations of this field have well understood all of these issues for a very long time. But, unfortunately, the widespread confusion concerning some of the most basic concepts of reversible computing has, in my opinion, held back the advancement of the field. It is high time that many more researchers let go of these previous misconceptions about what reversible computing means, and think outside of that earlier box.

**Today, finally,** with the end of scaling of conventional silicon CMOS in sight, the time is ripe for reversible computing to gain widespread attention. As mentioned earlier, the semiconductor industry is facing increasingly-insurmountable barriers to making continued progress along their conventional technology development path, barriers which threaten to bring progress to a near-halt in only roughly a decade. Developing any radically different alternative technology to the point where it can be economically viable will probably take at least that long, even with major new investments. So, we need to start now. And, crucially, *all* non-reversible approaches are ultimately dead ends, in the sense that, at best, they might get us one or a few technology generations beyond end-of-line silicon CMOS. But then, due to Landauer's principle,[10] they too will inevitably run out of steam. Even the other radical new concepts being explored, such as analog or spike-based neural computing, will eventually reach a limit, if they are not designed to *also be reversible*, since the fundamental laws of thermodynamics always hold for any physical system, regardless of whether we happen to view that system as digital or analog.[62] And finally, even if such very far-reaching concepts as quantum computing are successful, they will only help to significantly speed up certain highly specialized classes of computations.[16,17] Reversible computing is the *only* completely general approach that could make *all* possible computations ever more energy-efficient over the long term.

Fortunately, governments around the world are now beginning to become aware of the fact that major new public initiatives on developing future computing technology will likely be required to help the computing industry to hurdle the looming roadblocks. This is necessary because nearly all of the R&D resources of the global semiconductor industry are currently tied up just in the effort to get through the last few semiconductor technology nodes, while surviving the ongoing, cost-driven consolidation of the industry, and so these firms cannot justify to their investors devoting much attention to developing risky new alternative technologies. For example, the Chinese government, perhaps realizing the high level of geostrategic importance that computing will have in this century, has recently started supporting the advancement of their domestic computing industry at high levels, and currently boasts the world's fastest (and most energy-efficient) supercomputer[63] as a result. This situation has created some concern in the US about the impact of global computing leadership on national competitiveness and security, and decision makers here are beginning to respond.

In 2015, the White House Office of Science and Technology Policy announced a Nanotechnology-Inspired Grand Challenge for Future Computing[64] that was focused on developing vastly more energy-efficient computing technology for machine learning applications. Meanwhile, the Intelligence Advanced Research Projects Activity (IARPA) is pushing towards the development[65] of a superconducting supercomputer,[66] and the U.S. Department of Energy's National Laboratories are cooperating with each other on an inter-lab "Big Idea" effort to help figure out how to develop more energy-efficient computing, beyond the level of the coming Exascale systems which are slated to be produced by 2021[67] (a newly-accelerated target date). And here at Sandia National Laboratories, in particular, strategic research efforts in the general area of "Beyond Moore Computing" have been gradually taking shape.

All this activity illustrates that there is currently no shortage of leadership initiative, at many levels, that is beginning to gather steam and to orient itself towards the challenge of fostering the next revolution in computing technology. But today, many of these efforts are still relatively unfocused, in the sense that many decision-makers have not yet perceived any clear path forwards for restoring long-term progress in general-purpose computing. But, clarity can be improved by realizing that in fact, there is indeed exactly one such path that is physically possible, namely reversible computing. Its enormous potential upside warrants significantly expanding total national-scale investments in future computing R&D in such a way as to aggressively pursue reversible computing, in concert with a broad portfolio of nearer-term efforts. It will be an absolute physical necessity for making any long-term progress, but an intensive new level of effort in the field will still be required to enable us to solve its significant engineering challenges.

Most crucially, new reversible device technologies are needed, since adiabatic CMOS is arguably too slow to offer significant overall near-term benefits for system-level cost-efficiency (we would require much cheaper low-leakage transistors). Research on developing (much faster) reversible superconducting circuits is still ongoing (see Box 3)—and some of these circuits have already been demonstrated[68] to dissipate less energy per device than the Landauer limit that applies to irreversible computing. This empirically validates the core theoretical argument for reversible computing. Meanwhile, a team[69] led by Ralph Merkle (a well-known pioneer in cryptography[70] and nanotechnology[71]) has designed improved versions of reversible nanomechanical logic[72,73] (See Box 4) that have been analyzed[74] to be over 100 billion times as energy-efficient as today's technology, while still switching on nanosecond timescales—far surpassing the energy-delay efficiency of any other technology that has been proposed to date. However, nanofabrication technologies that could build these kinds of atomically-precise devices still need to be developed. And in the meantime, other, nearer-term concepts for new reversible devices are needed. Generally speaking, there's a pressing need for physicists who are working on developing new device concepts to refocus their efforts on designing their new devices with reversible operation in mind, since that is the *only* way that any new device can possibly surpass the practical capabilities of end-of-line CMOS technology by many orders of magnitude, as opposed to, at most, only one or a few.

And, a final significant challenge is that very advanced new high-precision engineering methods will be needed in order to produce the very high-quality oscillators that would be required to drive any of the various synchronous adiabatic reversible architectures with the requisite level of energy efficiency. Or, one possible alternative that I have been exploring recently is a concept that I call Asynchronous Reversible Computing[75,76,77,78,79] (ARC), which is, more precisely, a quasi-asynchronous type of ballistic reversible logic that is intended to be less demanding, in terms of clocking requirements, than all previous, synchronous design concepts. And there's yet another, even more radically unconventional concept I've explored for reducing clocking requirements which I call chaotic logic,[80,81,82] but that one is still in its infancy.

**To be clear,** reversible computing is by no means easy. All of the engineering problems mentioned above are very challenging to solve. Achieving efficient reversible computing, in any technology, will likely require a fairly thorough overhaul of our entire chip design infrastructure, from the fabrication process to construct novel device structures, possibly using new materials, all the way up through at least the processor architecture level. We'll need new design tools, new hardware description languages, and new hardware designs at many levels, with new supporting software. We'll have to re-train a large part of the digital engineering workforce to use the new design methodologies. I would guess that the total cost of all of the new investments in education, research, and development that will be required in coming decades will most likely run well up into the billions of dollars. It's a future-computing moonshot.

But, the difficulty of these challenges would be, in my opinion, a very poor excuse for not facing up to them. To me, at the present moment, when we seem to have arrived at an historic junction point in the evolution of computing technology, the decision that humankind faces appears to be quite stark. Do we want to effectively just give up on the future of computing, and begin getting used to idea that our technology development will soon plateau and then stagnate within a self-imposed state of affairs in which we will never able to carry out very much more computation with any given supply of energy than we will already be able to do a mere decade or so from now?

Or, do we want to seize this opportunity, begin blazing this new trail with collective gusto, and thereby open the door to a newly-unbounded future in which, over time, we may become able of carrying out *indefinitely many orders of magnitude* more computation with any given supply of energy then we can do today, together with all the indescribably far-reaching possible consequences for the long-term future of our civilization that such a limitless new capability could enable?

One thing is certain: We can't hope to successfully engineer that far-greater future for ourselves until we are willing to acknowledge that, yes indeed, *it can be done*, and then focus a substantial part of our collective energies towards achieving that goal.

*Illustrations follow on subsequent pages.*

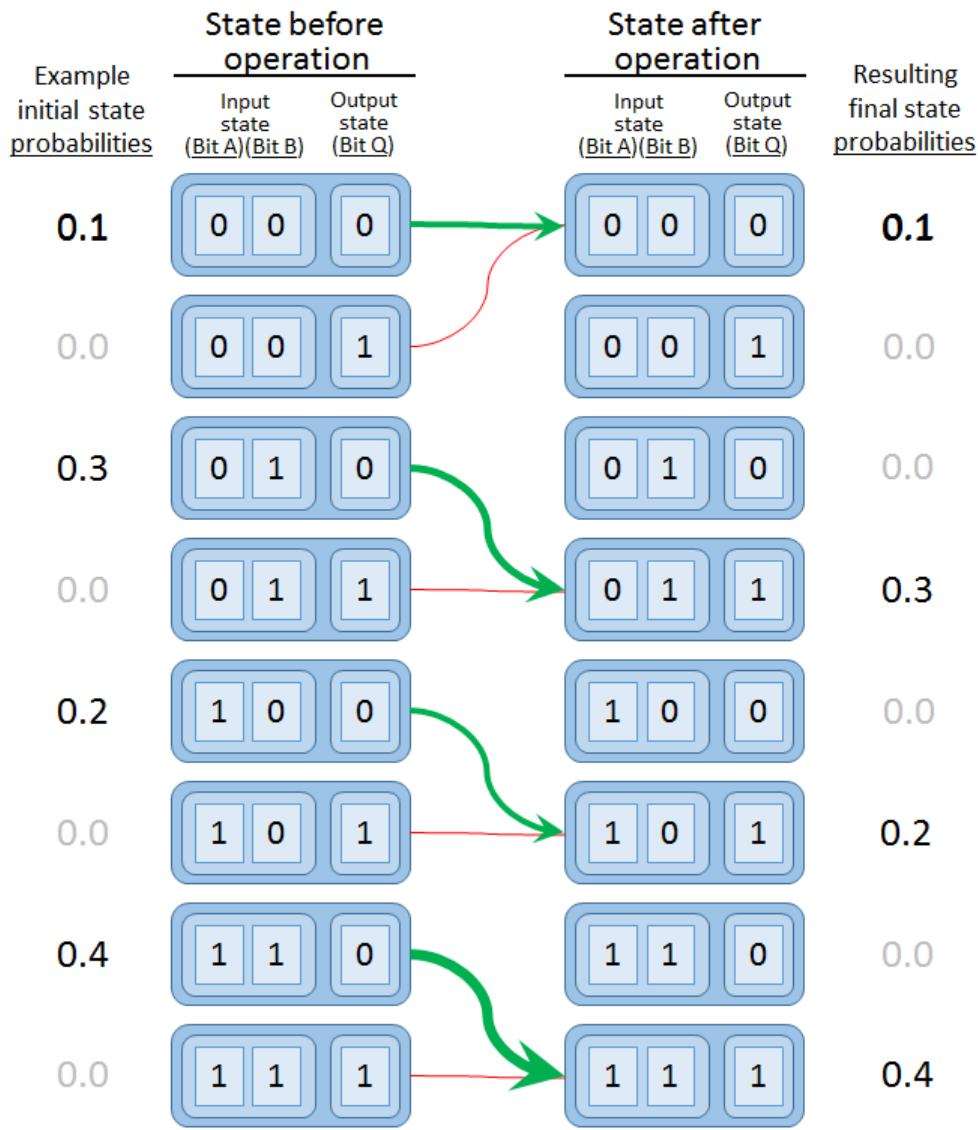

**Box 1.** An example of a conditioned reversible computational operation; this OR gate is logically reversible under the precondition that the output bit Q is initially 0. One can design simple mechanisms (see next box) that implement this computational operation in such a way that their physical operation approaches perfect thermodynamic reversibility (dissipating an amount of energy approaching zero), so long as the probability that the precondition is initially violated is zero, or approaches zero.[83,58] Here we show some arbitrarily-chosen initial state probabilities within an example operating context in which there is zero probability of violating the precondition. Note that none of the states bearing nonzero probabilities are merged together. This condition is perfectly sufficient, within a given operating context, for avoiding any expulsion of computational entropy, or energy dissipation due to Landauer's principle. Traditional models of reversible computing don't recognize this possibility, and so are overly restrictive. Claims such as "two-input Boolean gates that perform AND or OR operations are fundamentally irreversible" are therefore misleading. However, reversible computing still requires substantial reworking of digital logic designs even when conditionally-reversible Boolean gates are used, due to the unavoidable constraint that information can't be discarded.

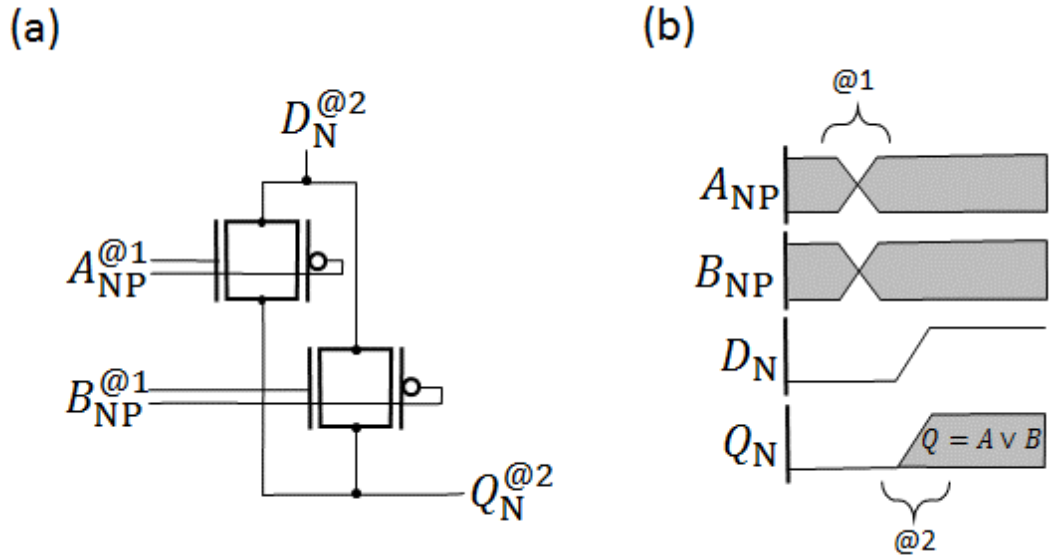

**Box 2.** (a) Example of a simple CMOS adiabatic circuit, in the 2LAL style, that implements the conditionally-reversible OR operation of the previous figure in an asymptotically reversible way (apart from transistor leakage). AND can be implemented similarly by reversing signal polarities. The circuit shown here consists of two CMOS transmission gates in parallel, each controlled by a dual-rail complementary (NP) pair of wires representing one of the logic inputs, $A$ or $B$. This circuit generates $Q_N$, the N-polarity (nFET-driving) version of the output, from the driving signal $D_N$; meanwhile, another copy of this circuit (not shown) driven by a complementary driver $D_P$ would generate the complementary output $Q_P$. (b) Sequence of operation. Initially $D$ and $Q$ are both low. The $A,B$ inputs transition adiabatically during time interval #1 ("at time 1" or "@1") to new valid levels; then, in a subsequent time interval @2, the driving control signal $D_N$ transitions high, and the output $Q_N$ follows it if and only if either input $A$ or $B$ is in the logic 1 state (with its N rail high, and P low). Therefore, $Q_N$ becomes the (N polarity half) of the logical OR of inputs $A$ and $B$, and similarly $Q_P$ becomes its complement. After computing $Q$, there are two reversible options for what happens next: (1) We can decompute the inputs $A,B$, reversibly restoring them to logic 0, and latching $Q$ in place, or (2) after downstream circuitry is finished utilizing the output $Q$, we can restore the driving signals $D_{NP}$ to their initial levels, thereby decomputing the value of $Q$, after which point the inputs $A,B$ can be safely transitioned to any new values. 2LAL was the first truly, fully adiabatic logic style capable of performing both logic and latching functions in the same structure, and this capability makes it useful for designing simple synchronous, pipelined reversible circuits. An animation[84] illustrates the structure and sequence of operation of a 2LAL shift register. The same operating principles can be applied in other domains (mechanical, etc.—see Box 4).

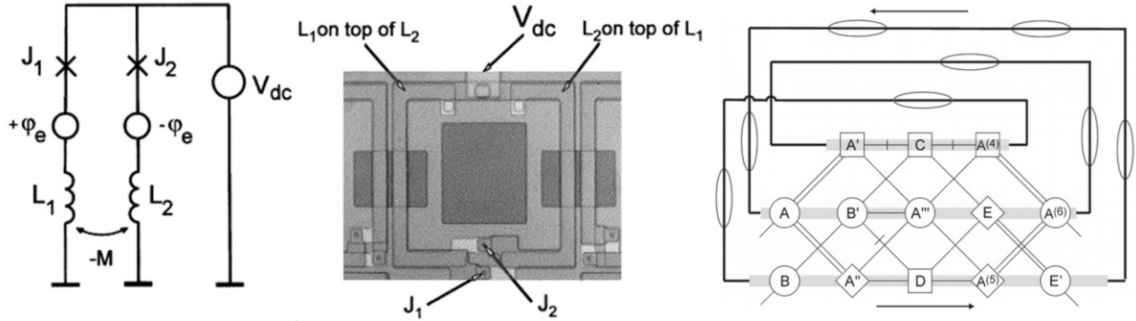

**Box 3.** Elements of a superconducting reversible logic scheme that has already been empirically demonstrated[68] by Vasili Semenov and colleagues to dissipate less energy per device than the Landauer minimum energy which would apply to logically irreversible devices. (Left) Circuit schematic for the basic element of this scheme,[85] which is a negative-mutual-inductance Superconducting Quantum Interference Device (nSQUID). (Middle) Micrograph of an nSQUID element as fabricated. (Right) Schematic of a circuit architecture for carrying out a simple reversible algorithm using these elements in a synchronous ballistic logic scheme clocked by Josephson vortices (ellipses) traveling along Long Josephson Junction (LJJ) transmission lines.

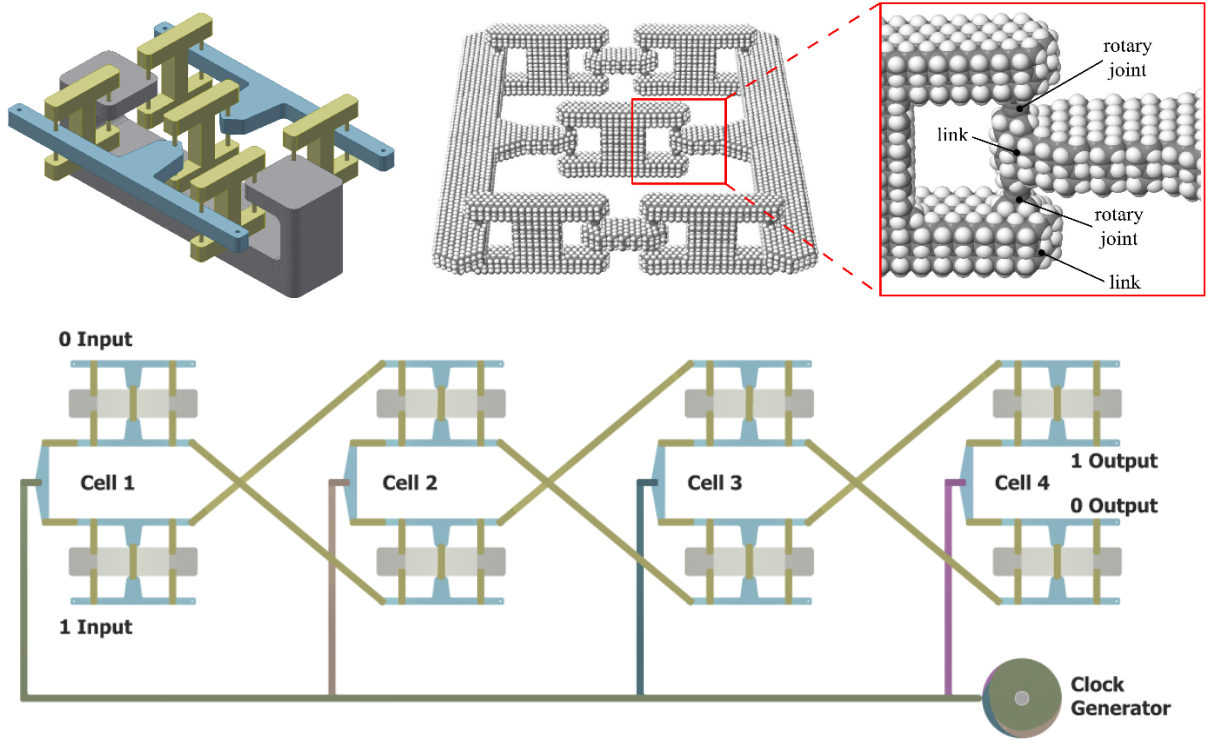

**Box 4.** The new nanomechanical rotary logic scheme of Merkle and colleagues.[69,73] This design envisions hydrogen-terminated nanostructured diamondoid material, and avoids all sliding contacts; the only bearings are rotary joints, which can be implemented using single carbyne bonds (upper right); these have a coefficient of rotary drag of less than $4\times10^{-35}$ J·s, according to detailed molecular dynamics simulations.[74] The basic structural element is the "lock" (upper middle & left), whose geometry ensures that only one of the blue bars at a time can be shifted out of its rest position. The choice of which bar is shifted can represent a bit. Synchronous reversible shift registers (bottom) can be designed, which can be viewed as essentially a mechanical-domain analog of a shift register in the 2LAL style of adiabatic CMOS (see animations[86,72]). With one more structure called a "balance," shown here driving shift register cells, universal reversible logic can be carried out. The authors calculate that a reversible NAND gate constructed with this approach would dissipate $3.9\times10^{-26}$ J, which is approximately 74,000× greater energy efficiency than any physically possible irreversible computer. Operating frequencies on the order of a GHz should be possible, and the authors estimate an aggregate power-performance of $1.28\times10^{12}$ GFLOPS (that is, 1.28 ZettaFLOPS!) per Watt of power dissipation. This illustrates the extreme power-performance benefits that could potentially be gained through reversible computing if the molecular manufacturing technologies required for this approach were developed.

> *For further reading*:
>
> - Popular science books that mention reversible computing (among other topics) include:
>   - *Engines of Creation: The Coming Era of Nanotechnology*, by K. Eric Drexler (1986).[87]
>   - *Minds, Machines, and the Multiverse: The Quest for the Quantum Computer*, by Julian Brown (2000).[88]
>   - *The Bit and the Pendulum: From Quantum Computing to M Theory—The New Physics of Information*, by Tom Siegfried (2000).[89]
> - The revised expanded manuscript edition of my MIT doctoral dissertation[90] includes various designs and analyses, and many references:
>   - *Reversibility for Efficient Computing*, by Michael P. Frank (1999).[91]
> - Technical books on reversible computing include:
>   - *Reversible Logic Synthesis*, by Anas N. Al-Rabadi (2004).[92]
>   - *Reversible Computing: Fundamentals, Quantum Computing, and Applications* by Alexis De Vos (2010).[93]
>   - *Towards a Design Flow for Reversible Logic*, by Wille and Drechsler (2010).[94]
>   - *Introduction to Reversible Computing*, by Kalyan S. Perumalla (2014).[95]
>   - *Theory of Reversible Computing*, by Kenichi Morita (2017).[96]
> - Additional books collecting a significant amount of relevant technical content:
>   - *Maxwell's Demon: Entropy, Information, Computing*, by Harvey Leff and Andrew Rex, eds. (1990).[97]
>   - *Feynman Lectures on Computation*, by Richard Feynman w. Anthony Hey and Robin Allen, eds. (1996).[98]
>   - *Feynman and Computation: Exploring the Limits of Computers*, by Anthony Hey, ed. (1999).[99]
>   - *Collision-Based Computing*, by Andrew Adamatzky, ed. (2002).[100]
>   - *Maxwell's Demon 2: Entropy, Classical and Quantum Information, Computing*, by Harvey Leff and Andrew Rex, eds. (2003).[101]
> - Conferences:
>   - In 2005, I organized the *First International Workshop on Reversible Computing*[102] as part of the ACM Computing Frontiers[103] conference in Ischia, Italy.
>   - Since 2009, there has been an annual series of workshops and conferences on *Reversible Computation*.
> - Technical Journals:
>   - The *IEEE Transactions on Emerging Topics in Computing*[105] is currently soliciting contributed articles for an upcoming special issue on *Design of Reversible Computing Systems*,[106] which I am co-editing.